*Article*# Super-operator Linear Equations and their Applications to Quantum Antennas and Quantum Light Scattering

Gregory Slepyan* and Amir Boag

School of Electrical Engineering, Tel Aviv University, Tel Aviv 69978, Israel

**\*** Correspondence: slepyan@post.tau.ac.il; Tel.: +972-54-737-89-17**Abstract:** In this paper we developed the resolvent method for super-operator equations with their applications in quantum optics. Our approach is based on the novel concept of linear super-operator acting on the Hilbert subspace of vector or scalar linear operators satisfying physically reasonable commutation relations. The super-operator equations for the electromagnetic (EM) field operators are formulated for the problems of quantum antenna emission and quantum light scattering by a dielectric body. The general solution of super-operator equation is presented in terms of the classical resolvent. In contrast to the classical case, it includes the ancillary components associated with the quantum noise even in the absence of absorption. The reason for it lies in the energy exchange between different spatial regions with various bases for the field presentation (it looks like losses or gain from the point of view of the correspondent region). A number of examples (two-element dipole antenna, plane dielectric layer, and dielectric cylinder with circular cross section) which demonstrate the physical mechanism of the appearance of noise are considered. It is shown, that antenna emission or scattering transforms the coherent properties of quantum light. This opens a new way of controlling the coherence in a direction dependent manner, a feature that can be useful in various applications of quantum technologies, including, quantum radars and lidars, and quantum antennas.

## 1. Introduction

The scattering of quantum electromagnetic (EM)-field is related to the wide research area stimulated by the potential applications in quantum computing and quantum informatics [1-4], quantum communications [4,5], quantum circuits [6], quantum antennas [7-13], quantum radars and lidars [14-16], sensing and measurements [17]. Currently this field is in the initial stages of its development. Probably, it will evolve along a path rather similar to classical electrodynamics and macroscopic antennas theory [18,19]. On the other hand, from time to time it is expected to encounter fundamental problems, whose origin is completely quantum and who don't have any classical analog (as one of such examples, one may note the manifestation of photons entanglement [1-5]). One of the first stages in the development of the field consists of exact analytical solutions of some model problems. This phase has started; the quantization of some exact analytical solutions of Maxwell equations obtained via the separation of variables (spherically layered multimode systems [20] and parabolic mirrors [21]) have been given.

As a parallel way of development, one can mention the high-frequency physical diffraction theories (for example, quantum diffraction by apertures [22,23]) and their comparison with the physics of simple quantum-optical devices (beam-splitters, interferometers, etc. [24]). You can expect that this phase will stimulate the establishment of universal numerical techniques, based on the so-called "discretization" of Maxwell equations (such as the method of characteristic modes [25-30], integral equations [31], finite

*Appl. Sci.* **2021**, *11*, x. https://doi.org/10.3390/xxxxx    www.mdpi.com/journal/applsci

differences [32], finite elements [33], etc). In the process of their making, the methods of classical electrodynamics will be an important element of the framework. As promising step in this direction, one may note the quantization of characteristic modes [34,35], the preliminary version of integral equations [36], a numerical approach based on the synthesis of the time-domain finite element analysis with the canonical quantization technique [37].

Further progress in the design of quantum devices often necessitates accurate solution of the problems of quantum field scattering. From a naïve point of view, the canonical field quantization looks like a formal replacement of classical fields by the corresponding operators, which act in the Hilbert space of wavefunctions (quantum states) and satisfy appropriate commutation relations [38,39]. However, the problem is not limited to the replacement of the classical values by the corresponding field operators. In classical electrodynamics, we note a widely used class of methods (including integral equations), in which the whole system is decomposed into a number of spatial regions. For example, for the scattering from a dielectric body, the integral equation is often formulated with respect to the field inside the body. The outside field is expressed through the inside one via an integral over the body volume. Another example is an antenna, defined as a device, which transforms the near field to the far field and vice versa. Such transformation is formally described by special integral relations based on the Green theorems [18]. In such cases, the quantum field the situation changes dramatically. The quantum field alongside with the field operators (which satisfy Maxwell equations) is characterized by the quantum state ($\Psi$-function). The introduction of partial regions requires the decomposition of the wavefunction of the whole space into those of the partial regions. The observable values are defined as expectation values of the corresponding operators with respect to the quantum state of the system [38,39]. The field operators which are related to the given partial region act on the correspondent partial wavefunction. As a result, the scattering of quantum EM-field will be accompanied by the appearance of entanglement in the new basis.

The formal description in our paper will be given in terms of super-operators. The super-operator $\bar{\mathbf{A}}$ is defined as a mapping of one area of quantum operators onto another one (in contrast to an operator, which defines the correspondence between two areas of functions). The formation of the mathematical theory of superoperators [40] began simultaneously with the emergence of quantum mechanics (von Neumann equation and Lindblad master equation [41] are related to the typical class of superoperator equations for matrix of density). In this paper, we introduce another concept of linear super-operator, which allows applying the integral equations theory to the scattering of quantum light and quantum antennas.

It is important to make a note on the role of losses in our concept of super-operator. The losses in the conventional quantum language mean the exchange of energy between the considered system ("small" system) and large external system described as a bath [42]. The "small" system is considered from the point of view of an open system. If we speak about thermal losses, the energy exchange means the energy transformation to heat. In this case, the bath takes the form of thermal reservoir. Such exchange is accompanied by the production of noise, without which the correct commutation relations become unattainable. A similar situation takes place for an active medium: the gain is necessarily accompanied by the noise [43-45].

The modeling of quantum light scattering and quantum antenna emission in this general language leads to the description of every partial region as an open quantum system. It continues the concept formulated previously for the quantum antennas [7]: the emission of quantum antenna looks like a loss from its point of view (in spite of the absence of any absorption). Therefore, the external space (with the receiver and the environment is considered as a bath of non-thermal origin). It allows applying the general strongly developed theory of open systems [42] to quantum antennas.

The subject of this paper is an application of the super-operator concept to the scattering of quantum light and quantum antennas. The paper is organized as follows: the general form of super-operator integral equation will be given in Section 2. The noise term will be considered and proved to have the form that guaranties the required form of commutation relations in general case. The application of the developed approach to the quantum antennas will be given in Section 3. The manifestations of the noise component in the transmission-reflection and scattering by the circular cylinder will be considered as examples in Sections 4 and 5, respectively. The conclusion and future perspectives are discussed in Section 6.

## 2. Super-operator integral equation for scattering of quantum light by a dielectric body

### 1.1. Preliminaries

The super-operator equation in our case may be written as

$$(\overline{\mathbf{I}} - \nu\overline{\mathbf{A}})\hat{x} = \hat{f}, \tag{1}$$

where $\nu$ is a spectral parameter (*c*-number), $\overline{\mathbf{I}}$ is the unit super-operator, $\hat{x}$ and $\hat{f}$ are operators in the space of quantum states ($\hat{x}$ is unknown, while $\hat{f}$ is given).

The formal solution of Equation (1) is $\hat{x} = \overline{\mathbf{R}}(\nu)\hat{f} = (\overline{\mathbf{I}} - \nu\overline{\mathbf{A}})^{-1}\hat{f}$, where $\overline{\mathbf{R}}(\nu)$ is the resolvent, which in our case is a super-operator too. As mentioned above, the operators $\hat{x}$ and $\hat{f}$ must satisfy the required commutation relations. Therefore, we encounter the main difference between the super-operator equations of the type (1) and their classical operator analogs. In contrast to the classical case, the resolvent $\overline{\mathbf{R}}(\nu)$ acts not on the whole Hilbert space, but on the subspace limited by the commutation relations. As will be shown below, the action of the resolvent is able to remove the operator $\hat{x}$ out of the defined subspace. As a result, one may obtain a mathematically exact, but physically incorrect solution with inadequate commutation properties. In other words, equation (1) is related to ill-posed mathematical problems and should be regularized for its correct application. Its regularization will be given in this paper. It is based on the introduction of the ancillary components, which are associated with a special type of quantum noise.

The configuration under consideration is shown in Figure 1. We decompose the operator of the longitudinal electric field into positive and negative frequency components [38,39] as $\hat{E}(\mathbf{x},t) = \hat{E}^+(\mathbf{x},t) + \hat{E}^-(\mathbf{x},t)$ and present it in the frequency domain as

$$\hat{E}^+(\mathbf{x},t) = i\int_0^\infty \hat{U}_k(\mathbf{x})e^{-i\omega_k t}dk, \tag{2}$$

where $\omega_k = kc$, $k$ is wavenumber, $c$ is light velocity in free space. A similar representation for $\hat{E}^-(\mathbf{x},t)$ is obtained from Equation (2) via its Hermitian conjugation. It will be useful to introduce the vector potential $\hat{A}(\mathbf{x},t) = \hat{A}^+(\mathbf{x},t) + \hat{A}^-(\mathbf{x},t)$ which is coupled with the field by the relation $\hat{E}(\mathbf{x},t) = -\partial_t \hat{A}(\mathbf{x},t)$. For the spectral density, we have Helmholtz equation

$$\nabla_\perp^2 \hat{U}_k + k^2 \varepsilon(\mathbf{x})\hat{U}_k = 0, \tag{3}$$

where symbol $\perp$ designates two-dimensional Laplacian and

$$\varepsilon(\mathbf{x}) = \begin{cases} \varepsilon, \mathbf{x} \in S \\ 1, \mathbf{x} \notin S. \end{cases} \tag{4}$$

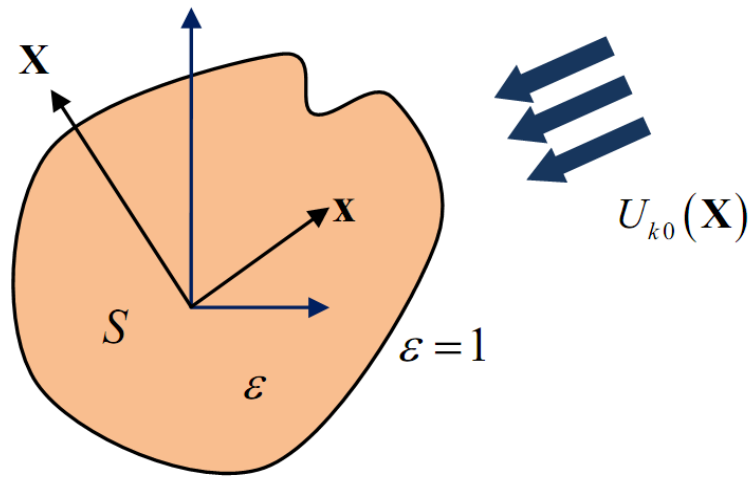

**Figure 1.** Configuration of the problem; two-dimensional dielectric cylinder with arbitrary cross-section $S$ is excited by the quantum light with spectral density $U_{k0}(\mathbf{X})$; $\mathbf{x} \in S$, $\mathbf{X} \notin S$.

The spectral density operator $\hat{U}_k(\mathbf{x})$ plays a role of the complex amplitude in the classical case. Like in the classical case, it satisfies boundary conditions at the boundary of the body (continuity of the tangential components of the field) and radiation condition. Following standard technique of classical electrodynamics, one can write the solution in terms of the resolvent. For the positive-frequency component of the field inside the body, the solution reads

$$\hat{U}_k(\mathbf{x}) = \hat{U}_{k0}(\mathbf{x}) - k^2(\varepsilon - 1)\int_S \Gamma(\mathbf{x}, \mathbf{x}'; k)\hat{U}_{k0}(\mathbf{x}')d\mathbf{x}', \tag{5}$$

where $\Gamma(\mathbf{x}, \mathbf{x}'; \nu)$ is the resolvent kernel defined by Equation (A5). The negative-frequency component $\hat{U}_k^+(\mathbf{x})$ inside the body may be obtained via Hermitian conjugation of Equation (5). It is generally accepted to write the resolvent as a function of the spectral parameter $\nu = -k^2(\varepsilon - 1)$ as in Equation (A1). We made in Equation (5) (and will use it in future) the transformation $\Gamma(\mathbf{x}, \mathbf{x}'; \nu) \to \Gamma(\mathbf{x}, \mathbf{x}'; k)$ because just wavenumber will take a role of a variable of integration in different relations.

If the inside field is known, the outside field may be found via the integration over the body volume

$$\hat{U}_k(\mathbf{X}) = \hat{U}_{k0}(\mathbf{X}) - k^2(\varepsilon - 1)\int_S g(\mathbf{X}, \mathbf{x}')\hat{U}_k(\mathbf{x}')d\mathbf{x}', \qquad (6)$$

where $\mathbf{X} \notin S$. For negative-frequency component, the equation may be found via Hermitian conjugation of Equation (6). Equations (5) and (6) look like the complete solution of the scattering problem: they satisfy Helmholtz equation over the whole space, boundary conditions and radiation conditions.

In contrast to the classical case, we need to add the commutation relation, which reflects the bosonic origin of EM-field. The commutation relation is local, thus it may be used in the ordinary form for homogeneous medium. As shown in Appendix B, it reads

$$\left[\hat{U}_k(\mathbf{x}), \hat{U}_{k'}^+(\mathbf{x}')\right] = \frac{\hbar k^2 c}{\pi \varepsilon(\mathbf{x})} \operatorname{Im}(g(\mathbf{x}, \mathbf{x}'))\delta(k - k') \qquad (7)$$

(In fact, it is obtained from relation for homogeneous dielectric space by replacing $\varepsilon \to \varepsilon(\mathbf{x})$).

*2.1. Commutation relation and noise component*

As one can see, it is impossible to consider Equations (5) and (6) as a solution for the field operators, because they don't satisfy the correct commutation rule (7). We will expand it next and make the suitable modifications, which lead to the physically correct solution. At first, let us explain the reason for such situation. To this end, we will calculate the commutator $\left[\hat{U}_k(\mathbf{x}), \hat{U}_{k'}^+(\mathbf{x}')\right]$ for operators $\hat{U}_k(\mathbf{x})$ and $\hat{U}_{k'}^+(\mathbf{x}')$ defined by relation (5) and its Hermitian conjugate. The final result is

$$\left[\hat{U}_k(\mathbf{x}), \hat{U}_{k'}^+(\mathbf{x}')\right] = \frac{\hbar k^2 c}{\pi \varepsilon} \operatorname{Im}(\Gamma(\mathbf{x}, \mathbf{x}'; k))\delta(k - k'). \qquad (8)$$

One can find the details of the calculations in Appendix B. As one can see, Equation (8) doesn't agree with the correct Equation (7), because of the appearance of the resolvent kernel, $\Gamma(\mathbf{x}, \mathbf{x}'; k)$, instead of the Green function, $g(\mathbf{x}, \mathbf{x}')$. This is a manifestation of scattering, which means that the classical resolvent technique cannot be directly transferred to the quantum case.

We are going now to their special regularization, whose main idea is the addition of the ancillary noise components. As a result, Equations (5) and (6) are modified in the following way:

$$\hat{U}_k(\mathbf{x}) = \hat{U}_{k0}(\mathbf{x}) - k^2(\varepsilon - 1)\int_S \Gamma(\mathbf{x}, \mathbf{x}'; k)\hat{U}_{k0}(\mathbf{x}')d\mathbf{x}' + \hat{F}_k(\mathbf{x}), \qquad (9)$$

where $\hat{F}_k(\mathbf{x})$ is the noise-field. For the negative-frequency component inside the body the solution may be obtained via Hermitian conjugation of Equation (9). The concept of noise looks similar to the lossy beam-splitter [44]; however, in our case their physical origin is different (the thermal losses are absent). We have for the noise-field $\langle\hat{F}_k(\mathbf{x})\rangle = \langle\hat{F}_k^+(\mathbf{x})\rangle = 0$ (the symbol $\hat{F}_k^+(\mathbf{x})$ here and in future denotes Hermitian conjugation of the noise field). The noise field operators commute with field operators $\hat{U}_{k0}(\mathbf{x}), \hat{U}_{k0}^+(\mathbf{x})$ and satisfy the commutation relation

$$\left[\hat{F}_k(\mathbf{x}), \hat{F}_{k'}^+(\mathbf{x}')\right] = -\frac{\hbar c k^2}{\pi \varepsilon} \delta(k - k') \text{Im}\left(\Gamma(\mathbf{x}, \mathbf{x}'; k) - g(\mathbf{x}, \mathbf{x}')\right). \tag{10}$$

The noise field may be presented at the form of series over the eigen-states $u_n(\mathbf{x})$ defined by Equation (A2):

$$\hat{F}_k(\mathbf{x}) = \sum_n \hat{f}_{kn} u_n(\mathbf{x}) \tag{11}$$

and similar for Hermitian conjugate operator. The pair $\hat{f}_{kn}, \hat{f}_{kn}^+$ is the creation-annihilation operators for noise mode with index $n$ at the given frequency $\omega = kc$. The commutation relations for them stem from Equation (10) and have the form

$$\left[\hat{f}_{kn}, \hat{f}_{k'n'}^+\right] = -\frac{\hbar c k^2 \nu}{\pi \varepsilon} \delta_{nn'} \delta(k - k') \text{Im}\left(\frac{1}{(\nu_n - \nu)\nu_n}\right). \tag{12}$$

*2.2. Observable values*

We have considered the quantum light scattering from the point of view of open quantum systems (for example, applied to the modeling of lossy beam splitter in [44]). The concept of composite quantum systems is used in it as a main framework. The composite quantum system is considered as a set of two (or more) subsystems, whose statistical properties completely define the behavior of the total system. For example, the two subsystems may be two atoms in molecule, two modes in EM-field, and two elements in electrical circuit, etc. In our case, the fields in two geometrical regions (the area inside the dielectric body and the whole outside space, near-field and far-field zones of antennas, etc.) play the role of subsystems. As a starting point, the Hilbert space of a composite system defined as a tensor product $\mathbb{S} = \mathbb{S}^{(1)} \otimes \mathbb{S}^{(2)}$ of Hilbert spaces of its subsystems $\mathbb{S}^{(1)}$ and $\mathbb{S}^{(2)}$ (superscripts 1 and 2 indicate the areas inside and outside the dielectric body, respectively) should be introduced.

The incident field is produced by the set of external sources distributed over the whole space. Such set may be considered as a composite quantum system and characterized by the density matrix, which is the operator in the state space $\mathbb{S}$. We will decompose the total source into two subsystems, in which the first one consists of sources outside the body and the second - of inside ones. We will assume, that the sources of different subsystems are uncorrelated. Therefore, the total density matrix takes the form of a tensor product of the density matrices of the partial regions: $\rho = \rho^{(1)} \otimes \rho^{(2)}$. Due to the noise component, the operators $\hat{U}_k(\mathbf{x}), \hat{U}_k^+(\mathbf{x})$ for the field inside the body act on the total space $\mathbb{S}$. For calculation of observable values, the field operators are conveniently present in the form of the tensor products. Operator (9) will have a form

$$\hat{U}_k(\mathbf{x}) = \left(\hat{U}_{k0}(\mathbf{x}) \otimes \hat{I}^{(2)}\right) - k^2(\varepsilon - 1) \int_S \Gamma(\mathbf{x}, \mathbf{x}'; k)\left(\hat{U}_{k0}(\mathbf{x}') \otimes \hat{I}^{(2)}\right) d\mathbf{x}' + \left(\hat{I}^{(1)} \otimes \hat{F}_k(\mathbf{x})\right), \tag{13}$$

and similar for the negative-frequency field $\hat{U}_k^+(\mathbf{x})$ (operators $\hat{I}^{(1,2)}$ denote the identity superoperators in the subspaces $\mathbb{S}^{(1,2)}$.

For an arbitrary operator $\hat{O}$ acting on the total space $\mathbb{S}$, the observable value is $\langle \hat{O} \rangle = Tr(\hat{O}\rho)$. Assuming that the fluctuations part is in vacuum state, we obtain for the observable field

$$\langle \hat{U}_k(\mathbf{x}) \rangle = \langle \hat{U}_{k0}(\mathbf{x}) \rangle - k^2(\varepsilon-1)\int_S \Gamma(\mathbf{x},\mathbf{x}';k)\langle \hat{U}_{k0}(\mathbf{x}') \rangle d\mathbf{x}' \qquad (14)$$

(with no support of fluctuations). For the intensity $\hat{O}(\mathbf{x}) = \hat{U}_k^+(\mathbf{x})\hat{U}_k(\mathbf{x})$, we have

$$\langle \hat{O}(\mathbf{x}) \rangle = \langle \hat{U}_k^+(\mathbf{x})\hat{U}_k(\mathbf{x}) \rangle = \langle \hat{U}_{k0}^+(\mathbf{x})\overline{\mathbf{R}}^+ | \overline{\mathbf{R}}\hat{U}_{k0}(\mathbf{x}) \rangle + \langle \hat{F}_k^+(\mathbf{x})\hat{F}_k(\mathbf{x}) \rangle_0, \qquad (15)$$

where $\overline{\mathbf{R}}(\cdot) = (\cdot) + \nu\int_S \Gamma(\mathbf{x},\mathbf{x}';k)(\cdot)d\mathbf{x}'$ is the resolvent operator, and $\langle ... \rangle_0$ denotes averaging over the vacuum state $|0\rangle\langle 0|$. As one can see from Equation (15), the second term describes the vacuum fluctuations of the intensity of the field inside the body due to the exchang of its energy with the outside space. It shows, that the scattering of quantum light is necessarily accompanied by the fluctuations. It looks similar to the relation between fluctuations and dissipation ("fluctuation-dissipative theorem" [42]). Therefore, one can speak here about "fluctuation-scattering theorem". Its origin is completely quantum: the second term in Equation (15) vanishes in the classical limit. Another origin of the noise component is the thermal radiation of the body environment. In this case, we are speaking about thermal bath instead of photonic one. Note that the statement of the quantum light scattering problem without accounting for any type of reservoir becomes physically contradictive.

## 3. Application to the quantum two-element antenna array

The purpose of this section is to present an application of the super-operator concept to a quantum two-element array of dipole emitters shown at Figure 2. This type of antennas is widely used in classical radio engineering and its application in super-resolving quantum radars was proposed in [15].

*3.1. Quantum noise and commutation relations*

The incident field (incoming modes) and emitted field (outgoing modes) may be described by the 4D columns of creation-annihilation operators $\{\hat{x}_1, \hat{x}_2, \hat{y}_1, \hat{y}_2\}$ and $\{\hat{x}_{1i}, \hat{x}_{2i}, \hat{y}_{1i}, \hat{y}_{2i}\}$, which are coupled via the resolvent matrix ($\overline{S}$-matrix) as

$$\begin{pmatrix} \hat{x}_1 \\ \hat{x}_2 \\ \hat{y}_1 \\ \hat{y}_2 \end{pmatrix} = \underbrace{\begin{pmatrix} r & t_1 & t_2 & t_3 \\ t_1 & r & t_3 & t_2 \\ t_2 & t_3 & r & t_1 \\ t_3 & t_2 & t_1 & r \end{pmatrix}}_{\text{Resolvent matrix}} \cdot \begin{pmatrix} \hat{x}_{1i} \\ \hat{x}_{2i} \\ \hat{y}_{1i} \\ \hat{y}_{2i} \end{pmatrix} \qquad (16)$$

Here $\hat{x}_{1,2}$ is the pair of outgoing mode creation operators at the first and second feeding transmission lines, $\hat{y}_{1,2}$ is the pair of outgoing creation operators at the first and second dipole emitters, the additional subscript *i* indicates the incoming mode at the corresponding channel. We assume that all channels are identical; therefore the reflection coefficient *r* is the same in all channels. A similar relation may be written for the annihilation operators via Hermitian conjugation of Equation (16).

The emitting properties of the antenna are characterized by the array factor [18]. For the quantum case it becomes the photonic operator. It's positive-frequency component in the *yz* plane is:

$$\text{AF}^+(\theta) = \hat{y}_1 e^{i\phi} + \hat{y}_2 e^{-i\phi}, \tag{17}$$

where $\phi = kd\cos\theta + \beta$, with $\theta$ being the polar angle in the spherical system with origin placed at the antenna geometrical center and $\beta$ is the phase shift. We assume that the system is lossless and reciprocal; thus, matrix $\bar{S}$ is unitary [46,47]. It leads to the coupling relations between its elements:

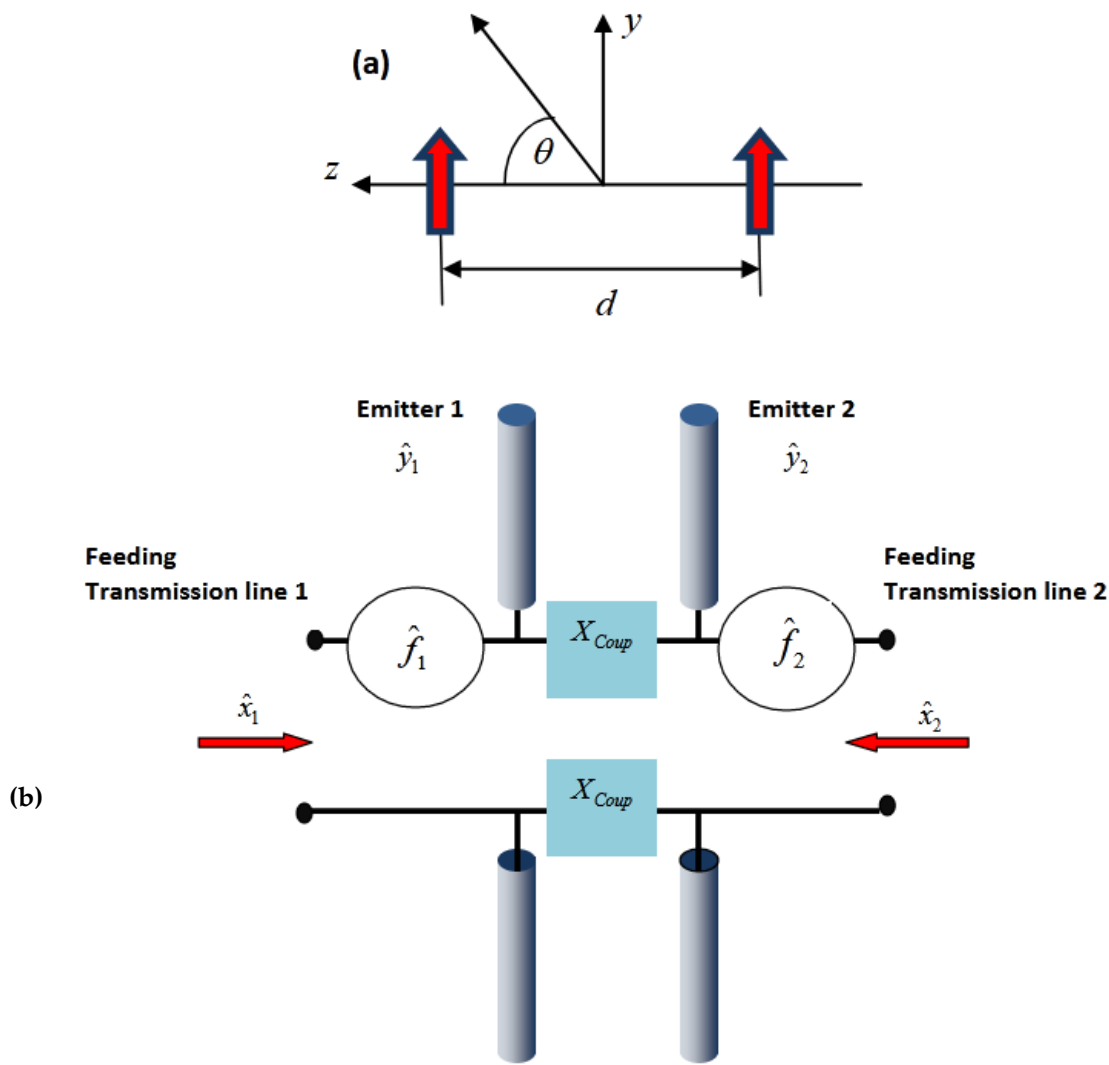

**Figure 2.** The scheme of two-element quantum antenna. (**a**) Configuration of antenna. Every emitting element of antenna is an electrical dipole. The dipoles are identical and identically oriented. (**b**) Equivalent electrical circuit of the antenna. The dipoles fed by the incident field via the corresponding transmission line (operators $\hat{x}_{1,2}$). The feeding is symmetrical (tap is placed at the central point of the dipole). Antenna elements are coupled via reactive impedance $X_{Coup}$ (no thermal losses at the antenna). A part of the incident power is reflected back; another part is distributed between the dipoles and emitted (operators $\hat{y}_{1,2}$). The feeding transmission lines are considered as an external environment. The elements coupling produces the special type of noise (operators $\hat{f}_{1,2}$).

$$|r|^2 + |t_1|^2 + |t_2|^2 + |t_3|^2 = 1, \tag{18}$$

$$\text{Re}(rt_1^* + t_2 t_3^*) = 0, \tag{19}$$

$$\text{Re}(rt_2^* + t_1 t_3^*) = 0, \tag{20}$$

$$\text{Re}(rt_3^* + t_1 t_2^*) = 0. \tag{21}$$

The operators satisfy the ordinary bosonic commutation relations $[\hat{x}_1, \hat{x}_1^+] = [\hat{x}_2, \hat{x}_2^+] = 1$, $[\hat{x}_{1i}, \hat{x}_{1i}^+] = [\hat{x}_{2i}, \hat{x}_{2i}^+] = 1$, $[\hat{y}_1, \hat{y}_1^+] = [\hat{y}_2, \hat{y}_2^+] = 1$, $[\hat{y}_{1i}, \hat{y}_{1i}^+] = [\hat{y}_{2i}, \hat{y}_{2i}^+] = 1$, and all other pairs of operators commute.

It is convenient for antenna applications to transform the equivalent 4-port network into a pair of coupled 2-port networks. This entails decomposition of the whole system into a pair of subsystems each of which is described in a different basis (actually, the decomposition of the whole antenna into a pair of dipole emitters and feeding transmission lines). To this end, Relation (16) may be rewritten as

$$\hat{\mathbf{x}} = \overline{A}\hat{\mathbf{x}}_i + \overline{B}\hat{\mathbf{y}}_i, \tag{22}$$

$$\hat{\mathbf{y}} = \overline{B}\hat{\mathbf{x}}_i + \overline{A}\hat{\mathbf{y}}_i, \tag{23}$$

where $\overline{A} = \begin{pmatrix} r & t_1 \\ t_1 & r \end{pmatrix}$ and $\overline{B} = \begin{pmatrix} t_2 & t_3 \\ t_3 & t_2 \end{pmatrix}$ are 2D blocks of the $\overline{S}$-matrix in Relation (16). Matrix $\overline{A}$ describes the coupling between emitters with their corresponding feeding lines, as well as mutual transmission line coupling. Matrix $\overline{B}$ describes the coupling of every emitter with the free space, as well as mutual coupling of emitters due to their near fields. The vectors $\hat{\mathbf{x}}, \hat{\mathbf{x}}_i, \hat{\mathbf{y}}, \hat{\mathbf{y}}_i$ are 2D columns of annihilation operators $\hat{\mathbf{x}} = \begin{pmatrix} \hat{x}_1 \\ \hat{x}_2 \end{pmatrix}$, $\hat{\mathbf{x}}_i = \begin{pmatrix} \hat{x}_{1i} \\ \hat{x}_{2i} \end{pmatrix}$, and similar for $\hat{\mathbf{y}}, \hat{\mathbf{y}}_i$.

For the classical field, we obtain Equations (22) and (23) for the corresponding classical values instead of operators. In our case, the incoming field appears only at the first and second channels. In this case, for classical field we should take $\hat{\mathbf{y}}_i = 0$ (it means, that the second subsystem is excited only via the coupling with the first one). However, such assumption becomes unphysical in the quantum case. The reason being the fundamental absence of empty channels in the quantum field. The non-excited channels in the quantum case should be considered as channels with zero number of photons (it means that $\langle \hat{\mathbf{y}}_i \rangle = 0$). However, it is impossible to ignore them because they are responsible for thermal and vacuum EM-fluctuations, omitting of which contradicts the correct commutation relations.

To facilitate the application of the resolvent theory, we should transform the system (22), (23) in the following way. Relation (23) is equivalent to

$$\hat{\mathbf{y}} = \bar{B}\hat{\mathbf{x}}_i + \hat{\mathbf{f}}, \tag{24}$$

where $\bar{A}\hat{\mathbf{y}}_i = \hat{\mathbf{f}}$. The first term in the right-hand side describes the excitation of the dipole emitters; while the second term corresponds to their coupling with the external space. From the point of view of the emitters it looks like energy is being lost, while it is not related to dissipation. Such "effective losses" may be positive and negative (the "effective gain" for receiving antenna). The operator $\hat{\mathbf{f}} = \begin{pmatrix} \hat{f}_1 \\ \hat{f}_2 \end{pmatrix}$ commutes with $\hat{\mathbf{x}}, \hat{\mathbf{x}}_i$, and $\langle \hat{\mathbf{f}} \rangle = 0$. The commutation relations for this operator stem from the commutation relations for $\hat{\mathbf{y}}_i$ and using Equations (18), (19) may be written as

$$\begin{aligned}
\left[\hat{f}_1, \hat{f}_1^+\right] &= \left[\hat{f}_2, \hat{f}_2^+\right] = 1 - |t_2|^2 - |t_3|^2, \\
\left[\hat{f}_1, \hat{f}_2^+\right] &= -t_2 t_3^* - t_3 t_2^*.
\end{aligned} \tag{25}$$

The non-zero right-hand sides in Equations (25) manifest the non-unitarity of the matrix block $\bar{B}$ (in spite of unitarity of the total $\bar{S}$-matrix in Relation (16)). The commutation Relations (25) completely coincide with the corresponding relations of lossy beam-splitter [44], where the right-hand sides are stipulated to produce the quantum noise due to dissipative processes. This is the reason to associate operator $\hat{\mathbf{f}}$ with the quantum noise too, in spite of the absence of any dissipation. It is a result of the separate description of the antenna emission and antenna feeding taking into account their interaction only at the final step. It is clear that the noise operator acts on the wavefunction of the second subsystem. It leads to the modification of the Thevenin-like equivalent scheme [18]: in contrast to the classical case, this creates the pair of additional noise sources (see Figure 2b).

*3.2. Transformation of the statistical properties of quantum light via the quantum antenna*

Let us consider the calculation of observable characteristics of quantum light. The antenna emission state is characterized by the wavefunction, which reads in the tensor form as

$$|\Psi\rangle = \sum_{n_1}\sum_{n_2} C_{n_1,n_2} |n_1 : \psi_1\rangle \otimes |n_2 : \psi_2\rangle = \sum_{n_1,n_2} C_{n_1,n_2} |n_1 n_2\rangle, \tag{26}$$

where $|n_\alpha : \psi_\alpha\rangle$ denotes a state with $n$ photons in mode number $\alpha = 1, 2$, $C_{n_1,n_2}$ are arbitrary coefficients, which satisfy the normalization condition. The directional dependence of the field properties is defined by the array factor given by Relation (17).

For our future purposes, it is convenient to move the directional dependence to the wavefunction and use the spatially independent array factor related to the antenna feed point. The situation is similar to the relation between the Schrodinger and Heisenberg pictures of quantum mechanics (with angle instead of time). Let us introduce the unitary operator

$$\hat{\mathbb{Z}}(\theta) = \sum_{m,n} e^{i(m-n)\phi} |mn\rangle\langle nm|, \tag{27}$$

where $\phi = (kd\cos\theta + \beta)/2$. The wavefunction of the field in the far zone may be introduced as

$$|\Psi\rangle_{rad} = \hat{\mathbb{Z}}(\theta)|\Psi\rangle = \sum_{mn} e^{i(m-n)\phi} C_{mn} |mn\rangle. \tag{28}$$

An arbitrary operator in such picture is transformed by the relation $\hat{\mathbb{F}} \to \hat{\mathbb{Z}}^+ \hat{\mathbb{F}} \hat{\mathbb{Z}}$. In particular, the array factor will be transformed by the following relation

$$\mathrm{AF}^+(\theta) \to \mathrm{AF}^+ = \hat{\mathbb{Z}}^+(\theta) \cdot \mathrm{AF}^+(\theta) \cdot \hat{\mathbb{Z}}(\theta) = \hat{y}_1 + \hat{y}_2. \tag{29}$$

Let us turn now to the analysis of antenna coupled with the feeding channels, as shown at Figure 2(b). Here, we introduce the Hilbert space of a composite system defined as a tensor product $\mathbb{S} = \mathbb{S}^{(x)} \otimes \mathbb{S}^{(y)}$ of Hilbert spaces of its subsystems $\mathbb{S}^{(x)}$ and $\mathbb{S}^{(y)}$ (superscripts (*x*) and (*y*) are related to the feeding lines and emitting dipoles, respectively). We will introduce the pure quantum state of light in the form of the tensor product $|\Psi\rangle = |\Psi_x\rangle \otimes |\Psi_y\rangle$. The states $|\Psi_x\rangle, |\Psi_y\rangle$ are given by

$$|\Psi_x\rangle = \sum_{n_1}\sum_{n_2} C_{n_1,n_2} |n_1 : \psi_1\rangle \otimes |n_2 : \psi_2\rangle,$$
$$|\Psi_y\rangle = |0 : \psi_3\rangle \otimes |0 : \psi_4\rangle, \tag{30}$$

where $|n_\alpha : \psi_\alpha\rangle$ denotes the state with *n* photons in the mode number $\alpha = 1,2,3,4$, similar to Equation (26) and $C_{n_1,n_2}$ are arbitrary coefficients, which satisfy the normalization condition.

The operator $\hat{\mathbf{y}}$ defined by Relation (24) acts in the space $\mathbb{S}^{(y)}$. It may be represented as an operator in $\mathbb{S}$ using the form of tensor product. We have

$$\hat{\mathbf{y}} = \hat{I}_x \otimes \bar{B}\hat{\mathbf{x}}_i + \hat{\mathbf{f}} \otimes \hat{I}_y, \tag{31}$$

where $\hat{I}_{x,y}$ are identity operators in spaces $\mathbb{S}^{(x,y)}$. As an example, let us calculate the observable $\langle \hat{\mathbf{y}}^+ \hat{\mathbf{y}} \rangle_{\mathbb{S}}$, which may be associated with an analog of intensity for the scattering by a dielectric body considered in Section 3 (the symbol $\langle ... \rangle_{\mathbb{S}}$ means the averaging over the state $|\Psi\rangle = |\Psi_x\rangle \otimes |\Psi_y\rangle$). Taking into account that for the noise in the vacuum state $\langle \hat{\mathbf{f}} \rangle = 0$ and using Relations (30) and (31), we obtain.

$$\langle \hat{\mathbf{y}}^+ \hat{\mathbf{y}} \rangle_{\mathbb{S}} = \langle \Psi_y | \bar{B}^* \hat{\mathbf{x}}_i^+ \bar{B}\hat{\mathbf{x}}_i | \Psi_y \rangle + \langle 0 | \hat{\mathbf{f}}^+ \hat{\mathbf{f}} | 0 \rangle. \tag{32}$$

The second term in Relation (32) corresponds to the support of fluctuations. It may be transformed to $\langle 0 | \hat{\mathbf{f}}^+ \hat{\mathbf{f}} | 0 \rangle = |t_2|^2 + |t_3|^2 = 1 - |r|^2 - |t_1|^2$.

Let us consider an example of the antenna excitation by a single photon in each feeding line. The wave function of the incident field is $|\Psi\rangle = |\Psi_x\rangle \otimes |\Psi_y\rangle = |11\rangle \otimes |00\rangle$. Assuming $t_1 t_2 + r t_3 = 0$ and $t_1 t_3 + r t_2 = 0$, we obtain that antenna interaction with the feed lines transforms the wavefunction to

$$|\Psi\rangle = |\Psi_x\rangle \otimes |00\rangle + |00\rangle \otimes |\Psi_y\rangle, \quad (33)$$

where

$$|\Psi_x\rangle = \sqrt{2}(rt_1)^*(|20\rangle + |02\rangle) + (r^2 + t_1^2)^*|11\rangle, \quad (34)$$

$$|\Psi_y\rangle = \sqrt{2}(t_2 t_3)^*(|20\rangle + |02\rangle) + (t_2^2 + t_3^2)^*|11\rangle. \quad (35)$$

As one can see from Equations (33)-(35), the feeding of the quantum antenna leads to an entanglement of quantum states of the external sources and the emitting dipoles. Following Equation (28) for transformation of the wavefunction (35) to the radiative form, we obtain

$$|\Psi_y\rangle_{rad} = \sqrt{2}(t_2 t_3)^*(|20\rangle e^{2i\phi} + |02\rangle e^{-2i\phi}) + (t_2^2 + t_3^2)^*|11\rangle. \quad (36)$$

Let us consider the special example where, we assume that antenna is perfectly matched to the feed lines ($r = 0$) and the feed lines are uncoupled ($t_1 = 0$). In this case, the entanglement between the antenna emission and antenna feeding becomes suppressed, in spite of the entanglement between the two emitters. As a result, we have for the wavefunction

$$|\Psi\rangle_{rad} = \frac{i}{\sqrt{2}}|00\rangle \otimes (|20\rangle e^{i\phi} + |02\rangle e^{-i\phi})_{rad}. \quad (37)$$

As one can see from Equation (37), in this special case two photons will be emitted with probability 0.5 from one of the emitters, while the probability to emit one photon from one of the emitters is equal to zero. It shows the fundamental property of quantum antennas: the process of antenna emission transforms the statistical properties of light (in particular, creates the inter-mode entanglement in the far field zone).

The angular distribution of the emitted field is characterized by the first and second-order correlation functions at the same point in the far zone. The corresponding Expressions for the quantum state (37) read

$$G^{(1)}(\theta) = \sin^2\theta({}_{rad}\langle\Psi|AF^+ \cdot AF|\Psi\rangle_{rad}) = 2\sin^2\theta, \quad (38)$$

$$G^{(2)}(\theta) = \sin^4\theta({}_{rad}\langle\Psi|AF^+ \cdot AF^+ \cdot AF \cdot AF|\Psi\rangle_{rad}) = 4\sin^4\theta\cos^2\phi = \\ 4\sin^4\theta\cos^2\left(\frac{1}{2}(kd\cos\theta + \beta)\right). \quad (39)$$

The first-order coherence is equal to the intensity of the emitted field. Its directive properties are equal to those of a Hertzian dipole. This results from the summation of partial powers emitted by each emitter without any interference mechanism for the quantum state (37). In contrast, the second-order coherence exhibits the narrow lobe-like directive properties, as follows from Equation (39). The angular dependence of the second-order correlation exhibits a set of identical lobes separated by zeros. Such behavior was predicted for the first time for the $N$-element antenna array in another quan-tum state in [10]. The number of lobes increases and their width decreases with increasing $kd$. The direction of the lobes is controllable by varying the phase shift $\beta$. It opens the way of electrically angular scanning by means of quantum antennas. Such ability may be important for applications in quantum radars and lidars [15].

## 4. Planar dielectric layer: reflection-transmission

The problem geometry is shown at the Figure 3a. The system is excited by the external current $f(x)$ distributed inside the dielectric layer along the *x*-axis and vanishing outside the layer. Therefore, the field is dependent only on one variable, *x*. The equation for the total field becomes

$$\frac{d^2 U_k}{dx^2} + k^2 \varepsilon(x) U_k = f(x), \tag{40}$$

where the permittivity $\varepsilon(x) = \varepsilon$ inside the layer and $\varepsilon(x) = 1$ outside it. The boundary conditions for the unknown field stem from the boundary conditions for the tangential component of the electric field at the boundaries of the dielectric layer [19,46]. The problem should be supplemented by the radiation condition at $x<0$ and $x>l$.

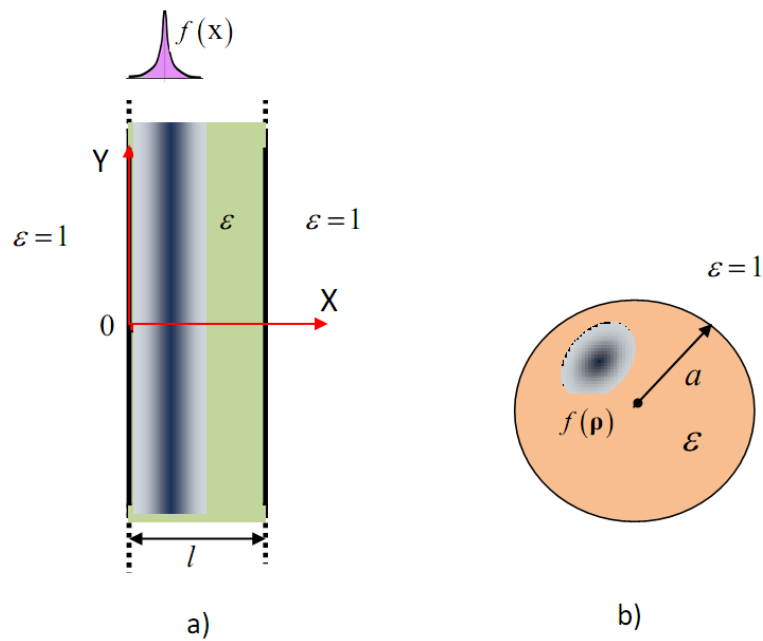

**Figure 3.** Examples of the dielectric structures for resolvent analysis: (**a**) dielectric layer; (**b**) dielectric cylinder of circular cross section; the sources (colored grey) placed inside the dielectric body.

The total field may be represented as a sum of incident and reflected-transmitted fields: $U_k = U_{k0} + \tilde{U}_k$. The incident field $U_{k0}$ is produced by the same source in the absence of the dielectric layer. Therefore, it satisfies equation $\frac{d^2 U_{k0}}{dx^2} + k^2 U_{k0} = f(x)$. For the reflected-transmitted field inside the layer, we have equation $\frac{d^2 \tilde{U}_k}{dx^2} + k^2 \varepsilon \tilde{U}_k = \nu U_{k0}$, where $\nu = -k^2(\varepsilon - 1)$, and $\frac{d^2 \tilde{U}_k}{dx^2} + k^2 \tilde{U}_k = 0$ outside. Our goal is to calculate the resolvent of the integral equation

$$\tilde{U}_k(x) = w_{k0}(x) + v\int_0^l g(x,x')\tilde{U}_k(x')dx', \qquad (41)$$

where $w_{k0}(x) = v\int_0^l g(x,x')U_{k0}(x')dx'$. It is an analog of the general integral Equation (A1) for the 1D case (where $g(x,x') = e^{ik|x-x'|}/2ik$ ).

We can represent the unknown field as

$$\tilde{U}_k(x) = \begin{cases} \dfrac{v}{2ik\sqrt{\varepsilon}}\int_0^l e^{ik\sqrt{\varepsilon}|x-x'|}U_{k0}(x')dx' + Ae^{ik\sqrt{\varepsilon}x} + Be^{-ik\sqrt{\varepsilon}x}, 0 \le x \le l, \\ Ce^{ikx}, x > l, \\ De^{-ikx}, x < 0, \end{cases} \qquad (42)$$

where *A,B,C, and D* are unknown constant coefficients. Using boundary conditions, we obtain the system of four linear equations

$$\begin{aligned} A + B - D &= -\frac{1}{\sqrt{\varepsilon}}f_+, \\ \sqrt{\varepsilon}(A-B) + D &= f_+, \\ Ae^{i\Phi_\varepsilon} + Be^{-i\Phi_\varepsilon} - Ce^{i\Phi} &= -\frac{e^{i\Phi_\varepsilon}}{\sqrt{\varepsilon}}f_-, \\ \sqrt{\varepsilon}\left(Ae^{i\Phi_\varepsilon} - Be^{-i\Phi_\varepsilon}\right) - Ce^{i\Phi} &= -e^{i\Phi_\varepsilon}f_-, \end{aligned} \qquad (43)$$

where $\Phi = kl, \Phi_\varepsilon = k\sqrt{\varepsilon}l$, $f_\pm = \dfrac{v}{2ik}\int_0^l e^{\pm ik\sqrt{\varepsilon}x'}U_{k0}(x')dx'$. System (43) is easily solvable. We need only two of unknowns, for which we have

$$\begin{aligned} A &= -\frac{\xi}{\sqrt{\varepsilon}}\left(f_+ e^{-2i\Phi_\varepsilon} - rf_-\right), \\ B &= \frac{\xi}{\sqrt{\varepsilon}}\left(rf_+ - f_-\right), \end{aligned} \qquad (44)$$

where $r = (1-\sqrt{\varepsilon})\cdot(1+\sqrt{\varepsilon})^{-1}$ is Fresnel coefficient for the normal incidence and $\xi = re^{2i\Phi_\varepsilon}\left(1 - r^2 e^{2i\Phi_\varepsilon}\right)^{-1}$.

Using solution (44) together with Relations (42), we obtain the solution inside the layer in the canonical form

$$\tilde{U}_k(x) = v\int_0^l \Gamma(x,x';k)U_{k0}(x')dx', \qquad (45)$$

where the function

$$\Gamma(x,x';k) = \frac{1}{2ik\sqrt{\varepsilon}} e^{ik\sqrt{\varepsilon}|x-x'|} +$$
$$\frac{1}{2ik\sqrt{\varepsilon}\left(1-r^2 e^{2ik\sqrt{\varepsilon}l}\right)} \left(-re^{ik\sqrt{\varepsilon}(x+x')} + r^2 e^{ik\sqrt{\varepsilon}(x-x'+2l)} + r^2 e^{-ik\sqrt{\varepsilon}(x-x'-2l)} - re^{-ik\sqrt{\varepsilon}(x+x'-2l)}\right) \quad (46)$$

may be associated with the resolvent kernel of integral operator and used for the analysis of commutation relation for the case of dielectric layer. In agreement with general Equation (A5), we have $\Gamma(x,x';-k) = \Gamma^*(x,x';k)$. Therefore, the commutation relation may be written as

$$\left[\hat{A}^+(x,t), \hat{E}^-(x',t)\right] = -\frac{i\hbar}{\pi\varepsilon} \int_{-\infty}^{\infty} \Gamma(x,x';k) k\, dk. \quad (47)$$

The support of the first term in Equation (46) gives the correct commutation relation. For the analysis of all other terms let us consider the analytical continuation to the complex *k*-plane. All these terms are analytical functions with the exception of the poles defined by the equation $r^2 e^{2ik\sqrt{\varepsilon}l} = 1$. All these poles are located in the lower half-plane. The functions are exponentially decreasing in the upper half-plane. Therefore, this component of the integral (47) may be closed in the upper half-plane and gives zero contribution to the commutator. Thus, the noise components of the fields disappear in agreement with [44].

### 5. Scattering by the dielectric cylinder

The purpose of this subsection is to find the resolvent for the dielectric cylinder of circular cross section. To this end, we will solve the Helmholtz equation:

$$\left(\frac{\partial^2}{\partial \rho^2} + \frac{1}{\rho}\frac{\partial}{\partial \rho} + \frac{1}{\rho^2}\frac{\partial^2}{\partial \varphi^2}\right) U_k + k^2 \varepsilon(\rho) U_k = f(\boldsymbol{\rho}) \quad (48)$$

with permittivity $\varepsilon(\rho) = \varepsilon$ inside the layer and $\varepsilon(\rho) = 1$ - outside. The right-hand-side in (48) is associated with the external force arbitrarily distributed inside the cylinder. The unknown function satisfies the boundary conditions for the longitudinal component of electric field at the boundary of the cylinder [19,47] and radiation condition outside.

The scattered field may be represented similarly to Equation (42) by

$$\tilde{U}_k(\boldsymbol{\rho}) = -\frac{i\nu}{4} \int_S H_0^{(1)}\left(k\sqrt{\varepsilon}|\boldsymbol{\rho}-\boldsymbol{\rho}'|\right) \tilde{U}_{k0}(\boldsymbol{\rho}') d\boldsymbol{\rho}' +$$
$$\nu \sum_n A_n J_n\left(k\sqrt{\varepsilon}\rho\right) e^{in\varphi}, 0 \leq \rho \leq a \quad (49)$$

and

$$\tilde{U}_k(\boldsymbol{\rho}) = \nu \sum_n B_n H_n^{(1)}(k\rho) e^{in\varphi}, \rho \geq a, \quad (50)$$

where $A_n, B_n$ are unknown coefficients, which should be founded from boundary conditions. We will omit here their calculations (see Appendix D).

The final result for the field inside the cylinder is

$$\tilde{U}_k(\boldsymbol{\rho}) = \nu \int_S \Gamma(\boldsymbol{\rho}, \boldsymbol{\rho}'; k) U_{k0}(\boldsymbol{\rho}') d\boldsymbol{\rho}', \tag{51}$$

where

$$\Gamma(\boldsymbol{\rho}, \boldsymbol{\rho}'; k) = -\frac{i}{4} H_0^{(1)}\left(k\sqrt{\varepsilon}|\boldsymbol{\rho}-\boldsymbol{\rho}'|\right) + \delta g(\boldsymbol{\rho}, \boldsymbol{\rho}'; k), \tag{52}$$

$$\delta g(\boldsymbol{\rho}, \boldsymbol{\rho}'; k) = \frac{i}{4} \sum_n W_n(k) J_n\left(k\sqrt{\varepsilon}\rho\right) J_n\left(k\sqrt{\varepsilon}\rho'\right) e^{in(\varphi-\varphi')}, \tag{53}$$

and

$$W_n(k) = \frac{H_n^{(1)\prime}(ka) H_n^{(1)}\left(k\sqrt{\varepsilon}a\right) - \sqrt{\varepsilon} H_n^{(1)}(ka) H_n^{(1)\prime}\left(k\sqrt{\varepsilon}a\right)}{H_n^{(1)\prime}(ka) J_n\left(k\sqrt{\varepsilon}a\right) - \sqrt{\varepsilon} H_n^{(1)}(ka) J_n'\left(k\sqrt{\varepsilon}a\right)}. \tag{54}$$

Relation (51) expresses the scattered field through the incident field inside the cylinder in the canonical integral form. Therefore, Relation (52) may be associated with the resolvent kernel of integral operator and used for the analysis of commutation Relation (18) and the existence of noise.

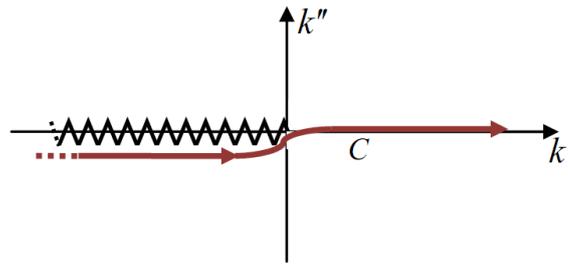

**Figure 4.** The contour $C$ (colored brown) is the integration contour in Equation (56). The cut is placed along the negative semi-axis and shown by the jagged line.

Let us transform Relation (18) for the dielectric cylinder to the form, similar Equation (47) for the 1D dielectric layer (integration over the whole axis instead of positive semiaxis with the subsequent integration in the complex plane). It is necessary to note, that here situation dramatically changes compared with the 1D case. The difficulty lies in the non-analyticity of the function $W_n(k)$ given by Relation (54) in the complex $k$-plane (this function is multi-valued with a branch point at $k=0$ in contrast to the dielectric layer case). One can see from well-known identities $J_{-n}(x) = (-1)^n J_n(x)$, $H_n^{(1)}(e^{j\pi}x) = -H_{-n}^{(2)}(x)$, and $H_n^{(2)}(e^{-j\pi}x) = -H_{-n}^{(1)}(x)$, that it is natural to place the cut in the complex $k$-plane along the negative semi-axis (jagged line in Figure 4). It easy to show, that $W_{-n}(x) = -W_n^*(e^{i\pi}x)$ and $W_n(k)$ goes to zero if $|k| \to \infty$ in the upper half-plane. Thus, for the quantity (53), we obtain the relation between two branches of complex variable $k$:

$$\delta g^*(\boldsymbol{\rho}, \boldsymbol{\rho}'; k) = \delta g(\boldsymbol{\rho}, \boldsymbol{\rho}'; e^{-i\pi}k). \tag{55}$$

The integration in Relation (47) for the cylinder may be transformed into an integral over contour $C$

$$\left[\hat{A}^+(\mathbf{x},t),\hat{E}^-(\mathbf{x}',t)\right] = -\frac{i\hbar}{\pi\varepsilon}\int_C \Gamma(\mathbf{x},\mathbf{x}';k)k\,dk \tag{56}$$

with its contour passing below branch-cut, as shown in Figure 4.

The contribution of the first term in Relation (52) to Relation (56) gives the conventional value of commutator without any noise components. However, in contrast to the 1D single-layer case (Section 4), the second term's contribution doesn't vanish. The reason being the impossibility to close the contour in the upper half-plane, where the function tends to zero (to do it one needs to cross the branch cut). This term is responsible for the noise current given by Equation (10).

## 6. Conclusion

In summary, we developed a novel method for the modeling of quantum light scattering and quantum antennas. It is formulated in terms of linear super-operator defined as an operator in the Hilbert space, which acts in the subspace of quantum state operators, satisfying the physically appropriate commutation relations. It is shown, that in the general case the naïve quantization via the formal exchange of the classical values by the corresponding operators leads to an ill-posed super-operator problem. The reason being, that the action of super-operator $\overline{\mathbf{A}}$ may throw the result outside of the required subspace, which violates the correct commutation relations. In order to satisfy them, one needs to do the regularization of super-operator equation.

The proposed regularization technique is based on the introduction of ancillary channels, which are associated with the novel type of quantum noise, whose origin is not related to any dissipation of energy. The regularized super-operator equation has the form $\overline{\mathbf{A}}\hat{E} = \hat{E}_0 + \hat{Q}$, where $\hat{Q}$ is a noise operator. Its form and commutation relations are determined in this paper. In general case, the super-operator equations may be solved numerically. Exact analytical solutions exist for special types of configurations. Sometime is promising the approximate analytical solution (for example, perturbation technique). The solution of the regularized super-operator equation may be expressed in terms of classical resolvent, which may be found by one of the methods of classical computational electromagnetics [48].

Simple applications of the theory are considered using several examples (two-element dipole quantum antenna, reflection-transmission in a 1D dielectric layer, and scattering by a dielectric cylinder). It is shown, that a special noise is produced via exchange of the energy between the body and the external space. From the "point of view" of the dielectric body, the process of multimode scattering looks like losses (or equivalent gain, if the energy enters from outside) in spite of the absence of any heating or real physical amplification. As a result, the exchange of energy is accompanied by the noise similar to that in systems with real physical dissipation. By the way, it becomes possible to speak about general "fluctuation-scattering theorem" similar to the well-known "fluctuation-dissipating theorem": the scattering of the quantum field is accompanied by fluctuations. In partial cases, a similar mechanism manifested itself in parallel-plane beam splitter [49] and time lenses [50]. In [49], the initial photon state being a product of number states in both modes is considered. It was shown that the lossless due to beam-splitter coupling create a noise in the photon number of each mode in spite of the total photon number being invariant and noise-free.

The theory developed in this paper may be generalized for many other scattering problems (different types of quantum antennas, scattering from 3D dielectric bodies, or dielectric bodies with different types of environment, etc.). It opens the way to using the scattering analysis as a tool of controlling the statistical properties of quantum light. It may be applied to various problems posed by the quantum technologies, metamaterials [51], radars and lidars [14,16], quantum circuits, and nanoantennas [52-55].

**Acknowledgments:** The authors acknowledge partial support by the H2020, project TERASSE 823878 and by the NATO Science for Peace and Security Programme under Grant id. G5860.

**Appendix A. Resolvent theory for classical operators**

The Fredholm integral equation of the second type in general form may be written as [56]

$$E(\mathbf{x}) - \nu \int_S g(\mathbf{x},\mathbf{x}') E(\mathbf{x}') d\mathbf{x}' = E_0(\mathbf{x}), \tag{A1}$$

where $E(\mathbf{x})$ is an unknown function, $\nu$ is a given parameter, while $E_0(\mathbf{x})$ (right-hand term) and $g(\mathbf{x},\mathbf{x}')$ (kernel) are a priori given functions. The eigen-functions $u_n(\mathbf{x})$ of integral Fredholm operator are defined by the homogeneous version of Equation (A1):

$$u_n(\mathbf{x}) = \nu_n \int_S g(\mathbf{x},\mathbf{x}') u_n(\mathbf{x}') d\mathbf{x}'. \tag{A2}$$

For the scattering problems the kernel of integral equation is the Green function of Helmholtz equation. For the two-dimensional (2D)-case $g(\mathbf{x},\mathbf{x}') = -\frac{i}{4} H_0^{(1)}(k|\mathbf{x}-\mathbf{x}'|)$ where $H_0^{(1)}(x)$ is the Hankel function of the zero order and the first type, As one can see, the integral operator kernel is symmetric: $g(\mathbf{x},\mathbf{x}') = g(\mathbf{x}',\mathbf{x})$, while non-Hermitian ($g(\mathbf{x},\mathbf{x}') \neq g^*(\mathbf{x}',\mathbf{x})$). Regarding the validity of the resolvent approach for this class of integral equations, in principle, the eigen-functions for non-Hermitian operators may not generate the complete basis system, and may not exist at all. However, the operator $\overline{\mathbf{A}}(\cdot) = \int_S g(\mathbf{x},\mathbf{x}')(\cdot) d\mathbf{x}'$ has some fundamental properties due to the special form of its kernel. First, the kernel has a weak singularity (integrable in the ordinary sense). Second, we can split the kernel $g(\mathbf{x},\mathbf{x}')$ into the real and imaginary parts and write $\overline{\mathbf{A}}(\cdot) = \overline{\mathbf{A}}_R(\cdot) + i\overline{\mathbf{A}}_I(\cdot)$, where subscripts *R, I* mean real and imaginary components, respectively. The singular component of operator $\overline{\mathbf{A}}(\cdot)$ is its real (Hermitian) part; while its imaginary part is regular. Thus, unbounded inverse $(\overline{\mathbf{I}} - \nu\overline{\mathbf{A}})_R^{-1}(\cdot)$ exists. Combining these two facts means that operator $\overline{\mathbf{A}}(\cdot)$ is a weak perturbation of Hermitian operator $\overline{\mathbf{A}}_R(\cdot)$ [57]. For weak perturbations of Hermitian operators, one can use the fundamental mathematical theory of pseudodifferential operators in Sobolev spaces [57] (in particular, the theorems on basis properties). For real *k* and $u(\mathbf{x}) \in \mathbf{L}^2(S)$, we have $\operatorname{Im} \int_S \int_S g(\mathbf{x},\mathbf{x}') u(\mathbf{x}) u^*(\mathbf{x}') d\mathbf{x} d\mathbf{x}' < \infty$ ($\mathbf{L}^2(S)$ is a space of Lebesgue squared integrable functions). In such case the operator $\overline{\mathbf{A}}(\cdot)$ is a bounded operator acting from $\mathbf{L}^2(S)$ to Sobolev space $\mathbf{H}^2(S)$. Such operator is completely continuous (compact) one. It means that the eigenvalues $v_n$ exist, nonzero, and lie in the upper half-plane ($\text{Im}(v_n) > 0$), and the resolvent set is nonempty [57].

It is easy to prove the orthogonality of eigen-functions in the following form

$$\int_S u_n(\mathbf{x}) \cdot u_m(\mathbf{x}) d\mathbf{x} = \delta_{mn} \tag{A3}$$

using the standard procedure. To that end, we should take Equation (A2) with index *m*, multiply it by $u_n(\mathbf{x})$, multiply (A2) by $u_m(\mathbf{x})$, subtract one from another, and integrate over *S*. The kernel may be represented as

$$g(\mathbf{x}, \mathbf{x}') = \sum_n \frac{1}{v_n} u_n(\mathbf{x}) \cdot u_n(\mathbf{x}'). \tag{A4}$$

The kernel

$$\Gamma(\mathbf{x}, \mathbf{x}'; v) = \sum_n \frac{1}{v_n - v} u_n(\mathbf{x}) \cdot u_n(\mathbf{x}') \tag{A5}$$

is referred as the resolvent kernel of integral Equation (A1) [58].

The kernel (A4) and resolvent kernel are coupled via Hilbert-Schmidt Relation [58]:

$$-g(\mathbf{x}, \mathbf{x}') + \Gamma(\mathbf{x}, \mathbf{x}'; v) = v \sum_n \frac{1}{(v - v_n) v_n} u_n(\mathbf{x}) \cdot u_n(\mathbf{x}'). \tag{A6}$$

The general solution of Equation (A1) may be efficiently presented in terms of the resolvent operator. It may be written as

$$E_v(\mathbf{x}) = E_{v0}(\mathbf{x}) + v \int_S \Gamma(\mathbf{x}, \mathbf{x}'; v) E_{v0}(\mathbf{x}') d\mathbf{x}'. \tag{A7}$$

One more identity will be useful in our future consideration:

$$\left(\frac{1}{v_n} - \frac{1}{v_m^*}\right) \int_S u_n(\mathbf{x}) \cdot u_m^*(\mathbf{x}) d\mathbf{x} = 2i \int_S \int_S \text{Im}(g(\mathbf{x}, \mathbf{x}')) u_n(\mathbf{x}) \cdot u_m^*(\mathbf{x}') d\mathbf{x} d\mathbf{x}'. \tag{A8}$$

It may be proven similarly to the ortogonality Relation (A3), however instead of Relation (A2) with index *m* its complex conjugated equation should be used.

**Appendix B: Quantization of free electromagnetic field in the cylindrical basis**

The purpose of this Appendix is the derivation of commutation Relation (7). We will consider the homogeneous space with permittivity $\varepsilon$ using the cylindrical basis. Let us present the positive-frequency component of the operator of vector potential in the form

$$\hat{A}^+(\mathbf{x}, t) = \int_0^\infty \frac{1}{\omega_k} \hat{U}_k(\mathbf{x}) e^{-i\omega_k t} dk. \tag{A9}$$

A similar relation may be written for the negative-frequency component of the electric field:

$$\hat{E}^-(\mathbf{x}, t) = -i \int_0^\infty \hat{U}_k^+(\mathbf{x}) e^{i\omega_k t} dk, \tag{A10}$$

where $\hat{U}_k(\mathbf{x})$, $\hat{U}_k^+(\mathbf{x})$ is the pair of Hermitially conjugate operators of spectral density. The problem is to find the commutation relation for the operators of spectral density, which gives the canonical bosonic commutation relation for the EM-field.

Let us present the operator of spectral density as a superposition of cylindrical waves:

$$\hat{U}_k(\mathbf{x}) = \sqrt{i}\tilde{E}_k \sum_{n=-\infty}^{\infty} \frac{1}{i^{-n}} J_n(k\rho) e^{-in\varphi} \hat{b}_{nk}, \quad (A11)$$

where $\tilde{E}_k = \sqrt{\hbar\omega_k k / 4\pi\varepsilon}$ is a normalization coefficient, $J_n(x)$ is Bessel function, and $\hat{b}_{nk}, \hat{b}_{nk}^+$ is a pair of creation-annihilation operators of cylindrical modes. We assume, that they satisfy the conventional bosonic-type commutation relation

$$\left[\hat{b}_{nk}, \hat{b}_{n'k'}^+\right] = \delta_{nn'}\delta(k-k'). \quad (A12)$$

Let us consider the commutator $\left[\hat{U}_k(\mathbf{x}), \hat{U}_{k'}^+(\mathbf{x}')\right]$. Substituting in it Relation (A11) and its Hermitian conjugate, we obtain

$$\left[\hat{U}_k(\mathbf{x}), \hat{U}_{k'}^+(\mathbf{x}')\right] =$$
$$-\frac{\hbar\sqrt{\omega_k\omega_{k'}kk'}}{4\pi\varepsilon} \sum_n \sum_{n'} i^{(n-n')} J_n(k\rho) J_{n'}(k'\rho') e^{i(n\varphi-n'\varphi')} \left[\hat{b}_{nk}, \hat{b}_{n'k'}^+\right]. \quad (A13)$$

Using Equation (A12), we transform it to

$$\left[\hat{U}_k(\mathbf{x}), \hat{U}_{k'}^+(\mathbf{x}')\right] = -\frac{\hbar\omega_k k}{4\pi\varepsilon} \sum_n J_n(k\rho) J_n(k\rho') e^{in(\varphi-\varphi')} \delta(k-k'). \quad (A14)$$

Using the addition theorem for cylindrical functions [59]

$$\sum_n J_n(k\rho) J_n(k\rho') e^{in(\varphi-\varphi')} = J_0(k|\mathbf{x}-\mathbf{x}'|), \quad (A15)$$

we obtain

$$\left[\hat{U}_k(\mathbf{x}), \hat{U}_{k'}^+(\mathbf{x}')\right] = -\frac{\hbar k^2 c}{4\pi\varepsilon} J_0(k|\mathbf{x}-\mathbf{x}'|) \delta(k-k'). \quad (A16)$$

Taking into account that $J_0(k|\mathbf{x}-\mathbf{x}'|) = -4\operatorname{Im}(g(\mathbf{x},\mathbf{x}'))$, we obtain commutation Relation (A16) in the final form

$$\left[\hat{U}_k(\mathbf{x}), \hat{U}_{k'}^+(\mathbf{x}')\right] = \frac{\hbar k^2 c}{\pi\varepsilon} \operatorname{Im}(g(\mathbf{x},\mathbf{x}')) \delta(k-k'). \quad (A17)$$

One can see, that commutation Relation (A17) really gives the conventional commutation relation for the EM-field operators, which is

$$-\varepsilon\left[\hat{A}(\mathbf{x},t), \hat{E}(\mathbf{x}',t)\right] = -\varepsilon\left[\hat{A}^+(\mathbf{x},t), \hat{E}^-(\mathbf{x}',t)\right] + c.c. = i\hbar\delta(\mathbf{x}-\mathbf{x}'). \quad (A18)$$

Substituting commutator $\left[\hat{A}^+(\mathbf{x},t), \hat{E}^-(\mathbf{x}',t)\right]$ in Relations (A9), (A10) and using Relation (A17), we obtain

$$\left[\hat{A}^+(\mathbf{x},t), \hat{E}^-(\mathbf{x}',t)\right] = -\frac{i\hbar}{\pi\varepsilon}\int_0^\infty \text{Im}(g(\mathbf{x},\mathbf{x}'))kdk. \tag{A19}$$

Thereafter, using the identity [59]

$$\int_0^\infty \text{Im}(g(\mathbf{x},\mathbf{x}'))kdk = \frac{\pi}{2}\delta(\mathbf{x}-\mathbf{x}'), \tag{A20}$$

we obtain Relation (A18).

**Appendix C: Derivation of Equation (8)**

Let us substitute the field representation (5) together with its Hermittian conjugate into the commutator of the operators. We have

$$\begin{aligned}\left[\hat{U}_k(\mathbf{x}), \hat{U}^+_{k'}(\mathbf{x}')\right] &= \frac{\hbar k^2 c}{\pi\varepsilon}\delta(k-k')[\text{Im}(g(\mathbf{x},\mathbf{x}')) + \\ &\quad \nu\int_S \Gamma(\mathbf{x},\boldsymbol{\rho};k)\text{Im}(g(\boldsymbol{\rho},\mathbf{x}'))d\boldsymbol{\rho} + \\ &\quad \nu\int_S \Gamma^*(\mathbf{x}',\boldsymbol{\rho};k)\text{Im}(g(\boldsymbol{\rho},\mathbf{x}))d\boldsymbol{\rho} + \\ &\quad \nu^2\int_S\int_S \Gamma(\mathbf{x},\boldsymbol{\rho};k)\Gamma^*(\mathbf{x}',\boldsymbol{\rho}';k)\text{Im}(g(\boldsymbol{\rho},\boldsymbol{\rho}'))d\boldsymbol{\rho}d\boldsymbol{\rho}'].\end{aligned} \tag{A21}$$

The second term in the right-hand-side of Equation (A21) may be transformed in the following way:

$$\begin{aligned}f_2(\mathbf{x},\mathbf{x}') &= \nu\int_S \Gamma(\mathbf{x},\boldsymbol{\rho};k)\text{Im}(g(\boldsymbol{\rho},\mathbf{x}'))d\boldsymbol{\rho} = \\ &\quad \frac{\nu}{2i}\left(\int_S \Gamma(\mathbf{x},\boldsymbol{\rho};k)g(\boldsymbol{\rho},\mathbf{x}')d\boldsymbol{\rho} - \int_S \Gamma(\mathbf{x},\boldsymbol{\rho};k)g^*(\boldsymbol{\rho},\mathbf{x}')d\boldsymbol{\rho}\right) = \\ &\quad \frac{\nu}{2i}\sum_m\sum_n\left(\frac{1}{(\nu_m-\nu)\nu_n}u_m(\mathbf{x})u_n(\mathbf{x}')\int_S u_m(\boldsymbol{\rho})u_n(\boldsymbol{\rho})d\boldsymbol{\rho} - \right.\\ &\quad \left.\frac{1}{(\nu_m-\nu)\nu_n^*}u_m(\mathbf{x})u_n^*(\mathbf{x}')\int_S u_m(\boldsymbol{\rho})u_n^*(\boldsymbol{\rho})d\boldsymbol{\rho}\right).\end{aligned} \tag{A22}$$

Using the orthogonality of the eigenmodes (A3) and Hilbert-Schmidt Relation (A8), we can rewrite Equation (A22) as

$$f_2(\mathbf{x},\mathbf{x}') = \frac{\nu}{2i}\left(\frac{1}{\nu}(\Gamma(\mathbf{x},\mathbf{x}';k) - g(\mathbf{x},\mathbf{x}')) - \sum_m\sum_n \beta_{mn}u_m(\mathbf{x})u_n^*(\mathbf{x}')\right), \tag{A23}$$

where

$$\beta_{mn} = \frac{1}{(\nu_m-\nu)\nu_n^*}\int_S u_m(\boldsymbol{\rho})u_n^*(\boldsymbol{\rho})d\boldsymbol{\rho}. \tag{A24}$$

For the third term in the right-hand-side of Equation (A21), we obtain by a similar argument

$$f_3(\mathbf{x},\mathbf{x}') = \nu \int_S \Gamma^*(\mathbf{x}',\boldsymbol{\rho};k) \operatorname{Im}(g(\boldsymbol{\rho},\mathbf{x})) d\boldsymbol{\rho} = f_2^*(\mathbf{x}',\mathbf{x}). \tag{A25}$$

For the last term in Equation (A21), using Equations (A22)-(A24), we obtain

$$\iint_{S\,S} \Gamma(\mathbf{x},\boldsymbol{\rho};k) \Gamma^*(\mathbf{x}',\boldsymbol{\rho}';k) \operatorname{Im}(g(\boldsymbol{\rho},\boldsymbol{\rho}')) d\boldsymbol{\rho} d\boldsymbol{\rho}' =$$

$$\sum_n \sum_m \frac{1}{(\nu_n - \nu)(\nu_m^* - \nu)} u_n(\mathbf{x}) u_m^*(\mathbf{x}') \iint_{S\,S} u_n(\boldsymbol{\rho}) u_m^*(\boldsymbol{\rho}') \operatorname{Im}(g(\boldsymbol{\rho},\boldsymbol{\rho}')) d\boldsymbol{\rho} d\boldsymbol{\rho}'. \tag{A26}$$

Using the representation of the Green function in the eigenmode basis (Equation (A4)), as well as their orthogonality (Equation (A3)), we transform Equation (A26) to

$$\iint_{S\,S} \Gamma(\mathbf{x},\boldsymbol{\rho};k) \Gamma^*(\mathbf{x}',\boldsymbol{\rho}';k) \operatorname{Im}(g(\boldsymbol{\rho},\boldsymbol{\rho}')) d\boldsymbol{\rho} d\boldsymbol{\rho}' =$$

$$\sum_n \sum_m \alpha_{mn} u_n(\mathbf{x}) u_m^*(\mathbf{x}'), \tag{A27}$$

where

$$\alpha_{mn} = \frac{1}{2i(\nu_n - \nu)(\nu_m^* - \nu)} \left(\frac{1}{\nu_n} - \frac{1}{\nu_m^*}\right) \int_S u_n(\boldsymbol{\rho}) u_m^*(\boldsymbol{\rho}) d\boldsymbol{\rho} \tag{A28}$$

with $\alpha_{mn} = -\alpha_{nm}^*$. Combining in Equation (A21) all obtained relations and using Hilbert-Schmidt Relation (A8), we obtain Equation (8) after some elementary transformations.

**Appendix D: Derivation of Equations (52),(53).**

The boundary conditions for the scattering field on the cylinder yield

$$\sum_n \left(A_n J_n(k\sqrt{\varepsilon}a) - B_n H_n^{(1)}(ka)\right) e^{in\varphi} = \frac{i}{4} \int_S H_0^{(1)}(k\sqrt{\varepsilon}|\mathbf{a}-\boldsymbol{\rho}'|) U_{k0}(\boldsymbol{\rho}') d\boldsymbol{\rho}', \tag{A29}$$

$$k\sum_n \left(A_n \sqrt{\varepsilon} J_n'(k\sqrt{\varepsilon}a) - B_n H_n^{(1)'}(ka)\right) e^{in\varphi} = \frac{i}{4} \int_S \frac{\partial}{\partial \rho} H_0^{(1)}(k\sqrt{\varepsilon}|\boldsymbol{\rho}-\boldsymbol{\rho}'|)\Big|_{\boldsymbol{\rho}\to\mathbf{a}} U_{k0}(\boldsymbol{\rho}') d\boldsymbol{\rho}', \tag{A30}$$

where $\mathbf{a} \in l$, $l$ is the contour of the cylinder, $|\mathbf{a}-\boldsymbol{\rho}'| = \sqrt{a^2 + \rho'^2 + 2a\rho'\cos(\varphi-\varphi')}$. For integration of the right hand sides in Equations (A29) and (A30), we use the addition theorem for cylindrical functions [60]

$$H_0^{(1)}\left(k\sqrt{\varepsilon}|\mathbf{a}-\boldsymbol{\rho}'|\right) = \begin{cases} \sum_n H_n^{(1)}\left(k\sqrt{\varepsilon}\rho'\right)J_n\left(k\sqrt{\varepsilon}\rho\right)e^{in(\varphi-\varphi')}, \rho \leq \rho' \\ \sum_n H_n^{(1)}\left(k\sqrt{\varepsilon}\rho\right)J_n\left(k\sqrt{\varepsilon}\rho'\right)e^{in(\varphi-\varphi')}, \rho \geq \rho'. \end{cases} \quad \text{(A31)}$$

Integrating over azimuthal dimension, we obtain

$$A_n J_n\left(k\sqrt{\varepsilon}a\right) - B_n H_n^{(1)}(ka) = \frac{i}{4} f_{nk} H_n^{(1)}\left(k\sqrt{\varepsilon}a\right), \quad \text{(A32)}$$

$$A_n \sqrt{\varepsilon} J_n'\left(k\sqrt{\varepsilon}a\right) - B_n H_n^{(1)'}(ka) = \frac{i}{4}\sqrt{\varepsilon} f_{nk} H_n^{(1)'}\left(k\sqrt{\varepsilon}a\right), \quad \text{(A33)}$$

where

$$f_{nk} = \int_S J_n\left(k\sqrt{\varepsilon}\rho'\right)e^{-in\varphi'} U_{k0}(\boldsymbol{\rho}') d\boldsymbol{\rho}'. \quad \text{(A34)}$$

Equations (A32), (A33) form a system of two linear algebraic equations with respect to the coefficients $A_n, B_n$. Finding $A_n, B_n$, substituting them to Equations (49), (50) and using the Wronskian of cylindrical functions, we obtain Equations (52), (53).